\documentclass[a4paper,12pt]{article}
\usepackage{epsfig}

\newcommand{\beq}{\begin{equation}}
\newcommand{\eeq}{\end{equation}}
\newcommand{\beqn}{\begin{eqnarray}}
\newcommand{\eeqn}{\end{eqnarray}}
\newcommand{\beqnn}{\begin{eqnarray*}}
\newcommand{\eeqnn}{\end{eqnarray*}}

\newcommand{\pT}[1]{p_{\perp_{#1}}}
\newcommand{\MT}{M_\perp}
\newcommand{\QT}{Q_\perp}

\newcommand{\REM}[1]{}

\begin{document}

\begin{center}
{\Large \bf Dileptons and Charm \& Bottom in Relativistic\\
  Heavy-Ion Collisions}

\vskip 5mm
{\sc K.~Gallmeister$^{1)}$, \underline{B.~K\"ampfer}$^{1)}$, 
O.P.~Pavlenko$^{1,2)}$}

\vskip 5mm

{\small
$^{1)}$ {Forschungszentrum Rossendorf, PF 510119, 01314 Dresden, Germany}
\\
$^{2)}$ {Institute for Theoretical Physics, 252143 Kiev - 143, Ukraine}
}
\end{center}

\vskip 5mm

\begin{center}
  \begin{minipage}{130mm}
    {\small
      We study the prospects to get information about the early and
      hot stages of deconfined matter produced in relativistic
      heavy-ion collisions by analyzing dilepton and single-lepton
      spectra. Energy losses of heavy quarks in deconfined matter and
      thermalization effects in hadron matter and their influence on
      lepton spectra are considered.}
  \end{minipage}
\end{center}

\vskip 10mm


\section{Introduction}

One can explore theoretically the properties of deconfined matter
by employing certain numerical methods in evaluating the theory of
strong interaction -- Quantum Chromo Dynamics (QCD) -- to get, for instance,
the equation of state \cite{lattice}. Phenomenological models can
supplement such analyses to arrive at a more qualitative understanding
of a system composed of quarks and gluons \cite{AP}.
Ultimately, however, one is interested in a verification whether such a 
state is realized in nature. It is usually thought that central collisions
of heavy ions at high energies offer the unique way to create transiently 
deconfined matter under laboratory conditions.  

There is a fairly long list of proposals of how to measure the properties
of such a novel matter state. 
Dileptons represent penetrating probes which are
considered since a long time to be a good messenger from the early
stages of the deconfined matter resulting in 
ultrarelativistic heavy-ion collisions.
Therefore, it should be possible to get direct information 
about the thermodynamical parameters of the hot and dense, 
strongly interacting system.
The problematic part about dileptons in heavy-ion collisions is that
there are quite a lot of different sources. Dalitz decays dominate
the low invariant mass region, while the high mass region is governed by
Drell-Yan (DY) dileptons. The preferable region for a thermal signal
from thermalized QCD deconfined matter
is the so-called intermediate mass region between the $\phi$ and 
the $J/\Psi$. In the resonance region below the $\phi$ the vector meson
decays provide a strong signal and also the thermal radiation 
from hadron matter can be probably best observed. 
In addition one has to be aware that, although most of heavier thermal
dileptons are produced in the very early stages of the hot matter, 
the production process continues during the whole evolution of the system
and only the space-time integrated yield can be measured. Therefore, some
efforts are needed to unfold a dilepton spectrum and to extract the
wanted information.

\begin{figure}[hbt]
  \begin{minipage}[t]{7cm}
    \psfig{file=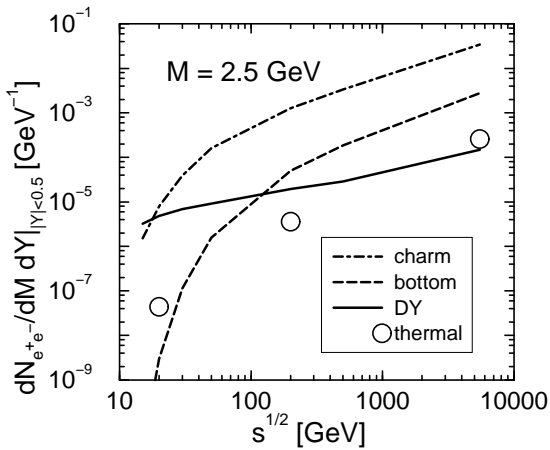,width=6cm,angle=-90}
  \end{minipage}
  \hspace*{7mm}
  \begin{minipage}[t]{7cm}
    \caption{\textit{The dependence of dilepton yields from the lowest-order 
        processes (DY and correlated semileptonic decays of open charm
        or bottom mesons) and the thermal source on $s^{1/2}$. Note that 
        only the thermal yield from deconfined matter is displayed.}} 
    \label{Fig:sqrts}
  \end{minipage}
\end{figure}

With increasing beam energies one generally expects higher matter 
temperatures and therefore a stronger signal of the thermal production.
At the same time, however, also other production channels for dileptons 
gain importance. In particular
the correlated semileptonic decays of open charm and bottom
mesons become strong sources.
To get an idea on the various competing sources, 
in fig.~\ref{Fig:sqrts} we compare the beam energy dependence of
the expected thermal signal with the DY yield and the dileptons
from correlated charm and bottom decays. 
(For the details about our modeling we refer the interested reader to 
\cite{GalKaePav98}. Here we mention that at large beam energies we estimate
the initial conditions of deconfined matter within the mini-jet model
\cite{mini-jet}. The DY process is calculated with standard procedures.
$c \bar c$ and $b \bar b$ pairs are produced in gluon fusion processes;
in lowest order the heavy quarks propagate back-to-back in the transverse
plane, and the hadronization into open charm and bottom mesons can be
approximated by a $\delta$ fragmentation scheme.) 

At SPS energies the charm and DY yields are of the
same order of magnitude, and the thermal signal is below both ones.
With increasing beam energy or $\sqrt{s}$ 
the thermal signal increases stronger than the DY yield.
On the other hand, the dilepton yields
from charm and bottom decays increase even stronger and result in
a background which is 
up to two orders of magnitude higher than the thermal signal.
Therefore it seems to be difficult to get thermal information from the
simple invariant mass spectrum. In what follows
 we are going to discuss whether one can
find such kinematical cuts which enable one to discriminate the
thermal signal from the background at very high beam energies such as
envisaged at RHIC and LHC (sect. 2). We also discuss the change of the 
heavy quark spectra by the deconfined medium and the resulting impact 
on the single lepton spectra stemming from charm and bottom decays
(sect. 3).
And finally we consider the influence of a dense hadron medium on the 
esfinal open charm spectrum at present SPS energies and show that the dilepton
spectra are correspondingly changed (sect. 4).   


\section{Perspectives for RHIC/LHC: dilepton spectra}

\begin{figure}[hbt]
  \begin{minipage}[t]{7cm}
    \psfig{file=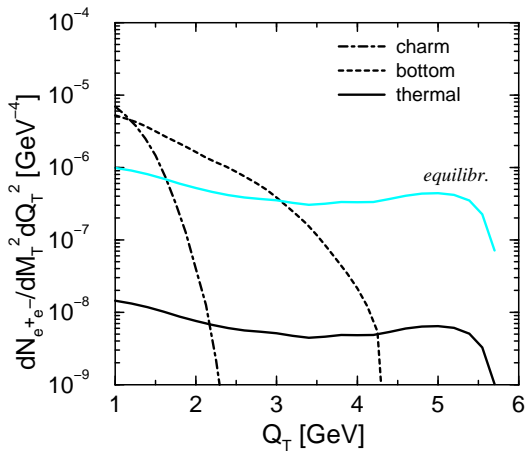,width=6cm,angle=-90}
  \end{minipage}
  \hspace*{7mm}
  \begin{minipage}[t]{7cm}
    \caption{\textit{The double differential rate for $\MT$ =
        5.25\dots5.75 GeV and $\pT{}> 2$ GeV at LHC within ALICE
        acceptance. The gray curve refers to a chemically
        equilibrated system, cf. 
        \protect\cite{GalKaePav98,GalKaePav99,GalKaePav98Erice} for
        details.}}
    \label{Fig:MTScal}
  \end{minipage}
\end{figure}

Since the kinematics of
heavy meson production and decay differs from that of thermal dileptons,
one can expect that special kinematical restrictions superimposed on
the detector acceptance will be useful for finding a window for
observing thermal dileptons in the intermediate mass continuum region.
As demonstrated recently \cite{GalKaePav98}, the measurement of the double
differential dilepton spectra as a function of the transverse pair
momentum $Q_\perp$ and transverse mass $M_\perp = \sqrt{M^2 + Q_\perp^2}$
within a narrow interval of $M_\perp$ offers a chance to observe
thermal dileptons at LHC.
The key observation here is to apply also a single--electron low--$\pT{}$
cut. Fig.~\ref{Fig:MTScal} shows the double differential spectrum for
a narrow interval of $\MT$ around 5.5 GeV and for $\pT{} > 2$ GeV.
One observes that these kinematical restrictions suppress the
background at large values of $\QT$, while 
the thermal signal obeys the so called $\MT$-scaling and extends
nearly up to the kinematical boundary.

\begin{figure}[hbt]
  \begin{minipage}[t]{7cm}
    \psfig{file=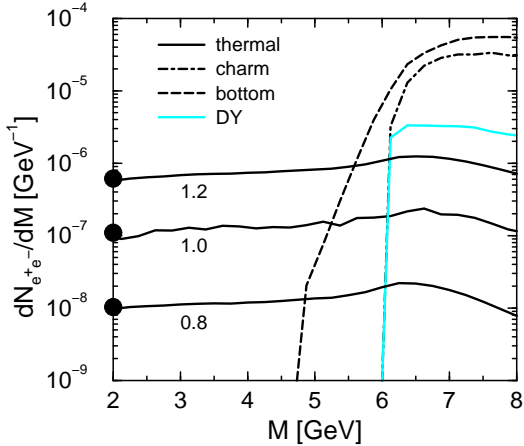,width=6cm,angle=-90}
  \end{minipage}
  \hspace*{7mm}
  \begin{minipage}[t]{7cm}
    
    \vspace*{-3mm}
    
    \caption{\textit{The invariant mass spectra of dileptons 
        within ALICE acceptance from
        the DY process, charm and bottom decays, and thermal
        emission. (The labels are for several initial temperatures
        in GeV.) 
        The single-electron low transverse momentum cut 
        is $p_\perp^{\rm min} =$ 3 GeV. The fat dots indicate the
        estimates of the low-$M$ thermal plateau as described 
        in \protect\cite{GalKaePav99}.}}
    \label{Fig:pTcut}
  \end{minipage}
\end{figure}

Another possibility to suppress the mentioned background processes is
to implement only a large enough low-$p_\perp$ cut on single
electrons \cite{GalKaePav99}. This opens a window for the thermal
signal in the invariant mass distribution.
Since the energy of individual decay electrons or positrons has a
maximum of about 0.88 (2.2) GeV in the rest frame of the decaying
$D$ ($B$) meson, one can expect to get a strong suppression of correlated
decay lepton pairs by choosing a high enough low-momentum cut
$p_\perp^{\rm min}$ on the individual leptons in the mid-rapidity region.
For thermal leptons stemming from deconfined matter there is no such upper
energy limit and for high temperature the thermal yield will not suffer
such a drastically suppression by the $p_\perp^{\rm min}$ cut as the
decay background.
The results of our lowest-order calculations of the invariant mass
spectrum with such a $p_\perp$-cut is displayed in
fig.~\ref{Fig:pTcut} again for LHC energies. 
One observes that 
the thermal dilepton signal with a single-electron low-momentum cut-off
$p_\perp^{\rm min} =$ 3 GeV exhibits an approximate plateau 
in the invariant mass region 2 GeV $\le M \le 2 p_\perp^{\rm min}$.

With both methods it is possible to extract the information about the
thermodynamical parameters of the very first stages of the deconfined
matter \cite{GalKaePav98Erice,KaePavGal98Erice}. 
But due to the quite low rates, it is questionable if these
cuts are experimentally feasible.

Recently, the ALICE-GSI group \cite{PBM} found that, 
via exact tracking and vertex reconstruction, one can suppress a substantial
part of the open charm and bottom decay electrons in the midrapidity
region. Therefore, the need of stringent cuts is relaxed somewhat and
realistic count rates are to be expected.
The announced heavy-ion programme of the CMS collaboration at LHC
\cite{CMS} looks also quite interesting as it can provide
complementary muon spectra in the midrapidity region.


\section{Energy losses of heavy quarks in deconfined matter} 

\begin{figure}[bt]
  \begin{minipage}[t]{7cm}
    \psfig{file=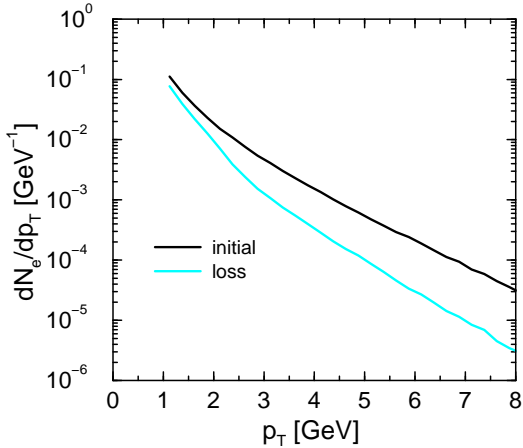,width=6cm,angle=-90}
  \end{minipage}
  \hspace*{7mm}
  \begin{minipage}[t]{7cm}
    \caption{\textit{The transverse momentum spectra of single electrons from
        D and B meson decays in the ALICE acceptance. Displayed are the
        spectra without (``initial'') and with energy loss according to
        model II in \protect\cite{GalKaePav98}. An additional
        vertex-cut is applied (see text).}}
    \label{Fig:singlepT}
  \end{minipage}
\end{figure}

Charm and bottom quarks traversing through deconfined matter
radiate gluons \cite{Zakharov} and lose energy. As a consequence
the transverse momenta of parent quarks and the resulting heavy
mesons and the emerging decay leptons are diminished.
Since the invariant mass of dileptons is
$M^2 = 2 p_\perp^+ p_\perp^- [\mbox{ch}(y^+ - y^-) - 
\cos (\phi^+ - \phi^-)]$, smaller values of the $p_\perp$'s cause
smaller values of $M$ 
($p_\perp^{\pm}$, $y^{\pm}$ and $\phi^{\pm}$ 
are the respective transverse momenta, and rapidities, and azimuthal
angles of leptons). 
Therefore, via energy losses the number of dileptons in the 
intermediate mass region is reduced. The effect is extensively studied
in \cite{PLB} and it turns out that, with realistic estimates of the 
energy loss, the background is not reduced below the thermal signal.
However, we would also like to point out that an explicit measurement
of the inclusive single-electron $p_\perp$-spectra 
from open charm and bottom decays contains valuable
information \cite{GalKaePav98}. 
Namely the energy losses sufficiently change the resulting
momentum distribution of the open charm and bottom mesons and, 
as a consequence, the decay electrons exhibit a significantly modified
$p_\perp$ spectrum (for details consult \cite{GalKaePav98}). 
Since such an effect does not appear in pp
collisions, the verification of a modified electron spectrum from identified
charm and bottom decays would offer a hint to the creation of deconfined
matter. 
The studies in \cite{PBM} demonstrate that tracking cuts offer
the chance to get a ''signal''-to-background ratio of 98\%, where
''signal'' means here the decay electrons from charm and bottom.
Therefore, such a measurement seems to be feasible with ALICE at LHC.
To illustrate the order of magnitude of the expected effect we show in
fig.~\ref{Fig:singlepT} the transverse momentum spectrum of decay
electrons from open charm and bottom mesons in a lowest-order
calculation as described above \cite{GalKaePav98}. 
Only electrons which seem to come from a point outside a sphere of 150 
$\mu$m around the primary vertex are counted.

Although charm and bottom are of the same order of magnitude, one can
fit the summed distribution by 
$dN_{e^-}/dp_\perp \propto \exp ( -p_\perp/T_e )$ in the interval
$p_\perp =$ 3\dots5 GeV and finds a change of the slope
parameter $T_e$ from 930 MeV (without energy loss) to
790 MeV (with energy loss). This difference is approximately the same
as could be expected in the PHENIX acceptance at RHIC.

One should stress that with such a measurement one can reveal the presence 
of a medium. Detailed information on thermodynamical state parameters
are, however, difficult to extract: a variation of the initial temperature
from 0.8 to 1.2 GeV causes only minor changes of the slopes of the
single-electron $p_\perp$ spectra.

\section{Dileptons in the intermediate mass region at CERN-SPS}

In present SPS experiments
($\sqrt{s} = 15 \cdots 20$ GeV) the CERES collaboration reports a dilepton
excess over the known hadronic cocktail in S + Au and Pb + Au collisions
at invariant masses $M < 1$ GeV \cite{CERES98}.
The order of magnitude of this excess
can be attributed to a thermal source stemming mainly from pion annihilation,
while the detailed shape of the spectrum is still matter of debate,
e.g.\ it might reflect an in-medium changed $\rho$ spectral function.
Similarly, in the intermediate mass region 
as accessible in the acceptances of the
HELIOS-3 and NA38/50 experimental set-ups, the conventional sources
Drell-Yan and open charm decays, known from pp collisions, seem also
not to account for the observed data, i.e. there is an excess too
\cite{excess,Scomparin}.  

\begin{figure}[hbt]
  \begin{minipage}[t]{7cm}
    \psfig{file=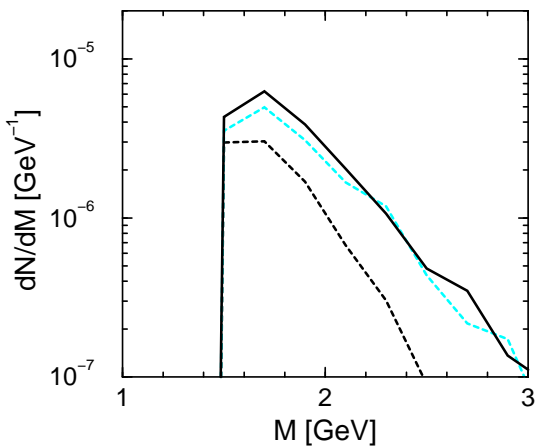,width=6cm,angle=-90}
  \end{minipage}
  \hspace*{7mm}
  \begin{minipage}[t]{7cm}

\vspace*{-4mm}

    \caption{\textit{Muon pair spectra with acceptance cuts of the
        NA50 apparatus. Full line: open charm mesons get a thermal 
        kick in their rest frame corresponding to a local temperature
        of 150 MeV; 
        the dashed lines are for primordial open charm without any
        medium effect. The gray line shows a change in the
        fragmentation scheme.}}
    \label{Fig:diffFrag}
\end{minipage}
\end{figure}

The NA50 data for central Pb-Pb collisions 
can be explained for an enhanced charm production \cite{Scomparin}.
A possible source could be a pronounced nuclear anti-shadowing of gluons
\cite{LinWang}. On the other hand, the charmed mesons can experience a
modification 
of their primordial spectrum via interactions with the dense hadron
medium before freezing out. As a working hypothesis one can
assume that the transverse open charm meson spectra in central heavy-ion
collisions at CERN-SPS look like the other hadron spectra.
Indeed, as shown in \cite{BK} the available transverse momentum spectra
of $\pi^{\pm}$, $K^{\pm}$, $K^0_s$, $p^{\pm}$, $\Lambda$, $\bar \Lambda$, $d$ 
at midrapidity can be described by a unique freeze-out
temperature of 120 MeV and a unique transverse flow with
averaged velocity of 0.41 c. Following a suggestion of \cite{LinWang}
one can give the charm mesons a randomly oriented thermal kick, which mimics
the thermalization process. As a consequence the resulting decay electron
spectrum is modified as displayed in fig.~\ref{Fig:diffFrag}.
One observes that, due to the very acceptance
of NA50 experiment, a change in the kinematical distributions of the
particles can lead to an apparent excess in the measured yield. 
This effect deserves further studies, e.g., by analyzing the transverse
dilepton spectra in the same mass region, which are presently prepared
by the NA38/50 collaboration \cite{NA38}. As a measure for the
theoretical uncertainties we also display in fig.~\ref{Fig:diffFrag} the
effect of changing the fragmentation scheme to Lund fragmentation with
Peterson function ($\epsilon=0.02$) and $m_c=1.5$ GeV.


\section{Summary}

It is obvious  that dileptons are very interesting and promising
signals if one wants to learn about the physics of highly excited,
strongly interacting matter. 
It is also clear that very much efforts have to be
put in to unfold the spectra to get the desired information.

We show that suitable kinematical cuts can suppress the background
dilepton yield. For a doubtless identification of the thermal dilepton 
signal, however, an explicit measurement of open charm and bottom would be 
preferable.
In addition, in the present contribution we report studies of two in-medium
effects: (i) the change of the charm \& bottom quark spectra by gluon radiation
in deconfined QCD matter and (ii) the change of the open charm meson 
spectra by thermalization in the hadron stage. Both effects are noticeable. 

\subsection*{Acknowledgments} 

We would like to thank Prof. I. Iori for the
organization of such a nice series of workshops and for the fruitful
and exciting scientific atmosphere in Bormio.
The work is supported by BMBF 06DR829/1.


\end{document}